\documentclass[a4paper,11pt]{article}

\usepackage{natbib}

\usepackage[dvips]{graphicx}
\usepackage{amsmath}

\begin{document}

\begin{center}
$~$\\
\vspace{1.5 cm}
{\LARGE SEA ICE BRIGHTNESS TEMPERATURE\\AS A FUNCTION OF ICE THICKNESS\\}
\vspace{0.2cm}
{\LARGE Part II: Computed curves\\for \\ thermodynamically modelled ice profiles\\}
\vspace{1.5cm}
{\Large Peter Mills\\}
\vspace{0.4 cm}
Peteysoft Foundation\\
1159 Meadowlane,
Cumberland ON,
K4C 1C3 Canada\\
\vspace{1.5cm}
\vspace{0.4 cm}
\vspace{0.2cm}
\includegraphics[width=0.75\textwidth]{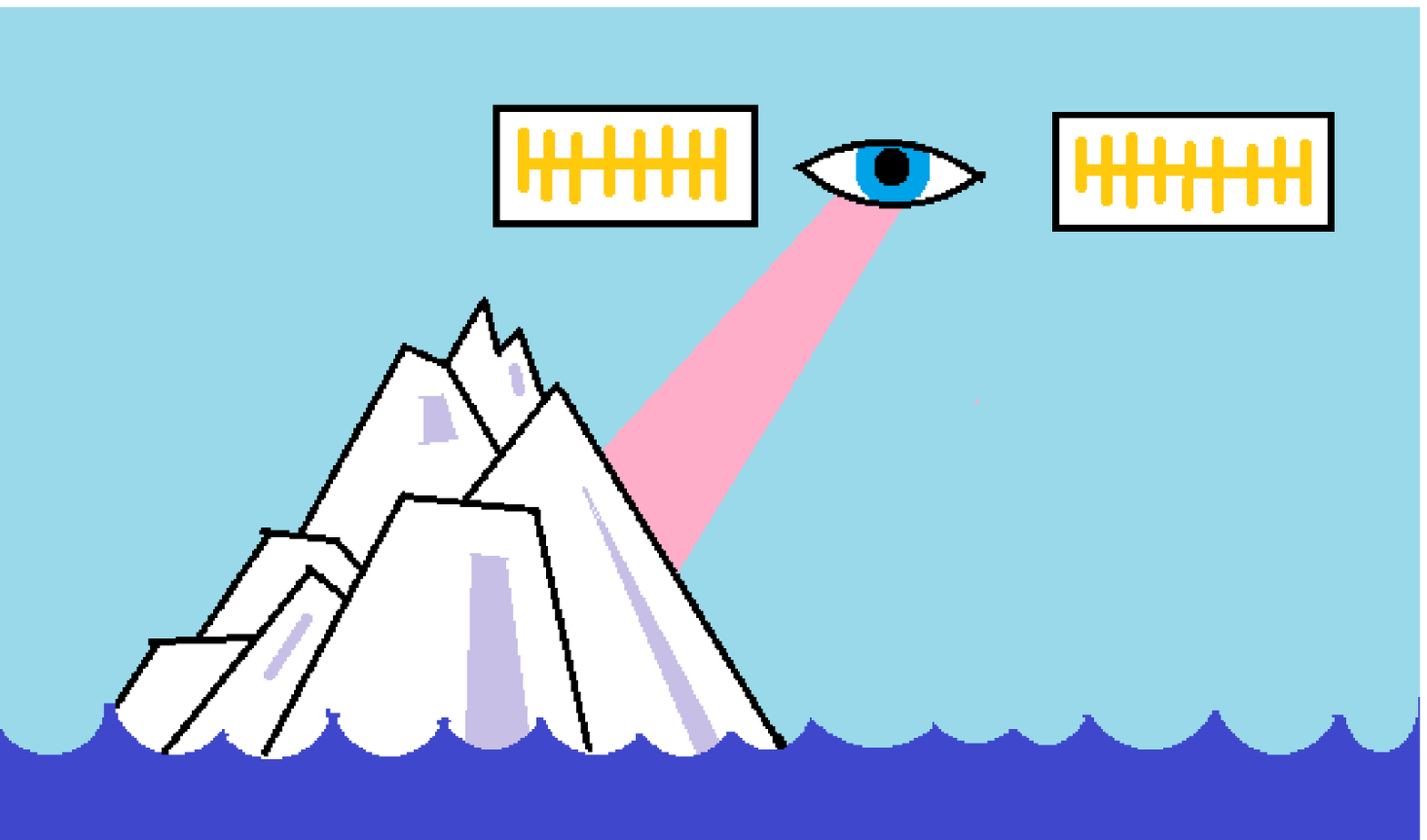}\\

\end{center}
\clearpage

\begin{abstract}
Ice thickness is an important variable for climate scientists
and is still difficult to accurately determine from microwave
radiometer measurements.
There has been some success detecting the thickness of thin ice
and with this in mind this study attempts to
model the thickness-radiance relation of sea ice at frequencies
employed by the Soil Moisture and Ocean Salinity (SMOS) radiometer
and the Advanced Microwave Scanning Radiometer (AMSR): between
1.4 and 89 GHz.
In the first part of the study, the salinity of the ice was determined
by a pair of empirical relationships, while the temperature was
determined by a thermodynamic model.
Because the thermodynamic model can be used as a simple ice growth model,
in this, second part, the salinities are determined by the growth model.
Because the model uses two, constant-weather scenarios representing
two extremes (``fall freeze-up'' and ``winter cold snap''), brine expulsion
is modelled with a single correction-step founded on mass conservation.
The growth model generates realistic salinity profiles, however it 
over-estimates the bulk salinity because gravity drainage 
is not accounted for.
The results suggest that the formation of ``skim'' on the ice surface
is important in determining the radiance signature of
thin ice, especially at lower frequencies, while scattering is
important mainly at higher frequencies but at all ice thicknesses.

\end{abstract}

\section{Introduction}

This study is a follow-up to a similar study \citep{Mills_Heygster2011c}
in which the relationship between sea ice brightness temperature at microwave
frequencies and ice thickness was examined.
As noted in the previous study, several other works
 have demonstrated a relationship between ice thickness
and SSM/I or AMSR-E brightness temperatures \citep{Martin_etal2004,
Naoki_etal2008,Kwok_etal2007,Hwang_etal2007} with at least one attempt to use
this relationship for the purpose of ice thickness retrieval \citep{Martin_etal2004}.
To understand the relationship, determine it's most likely functional form or forms
and to set definite bounds on it would be of great value to ice remote
sensing efforts.
In particular, satellite retrieval of sea ice thickness in all weather conditions
would be invaluable for climate scientists as it can help determine
a number of important flux variables in the arctic regions such as heat and
salt and fresh water.
At present, the thickness of thin ice can be determined fairly accurately
using infrared imagery, however this is quite cloud-sensitive, whereas
microwave imagery is not.

In the first part, parameterised empirical relationships were used to relate
bulk salinity to ice thickness and to determine the vertical salinity profile
within the ice.  In this second part, we will use a thermodynamic ice growth model, 
based primarily on first principles, to determine the salinity of the ice.

In Section \ref{ice_growth_model} we will describe in detail the thermodynamic
ice growth model, while in Section \ref{growth_results} we will present the
results of the that model. 
Finally, in Section \ref{results} we will apply the modelled profiles to the microwave emissivity simulation.
All other parameters, in particular the emissivity model set up,
will be the same as in the previous study and so will not be described.

\section{Ice growth model}
\label{ice_growth_model}

\begin{table}
\caption{Thermodynamic model parameters for the two weather scenarios}
\label{thermo_parm}
\begin{center}
\begin{tabular}{|l|cc|}
\hline
 & Scenario & \\
\hline
Parameter & Fall freeze-up & Winter cold snap \\
\hline\hline
Wind speed [m/s] & 2. & 10. \\
Air temperature [K] & 270. & 260. \\
Relative humidity & 0.5 & 0.1 \\
Cloud-cover & 0.5 & 0.1 \\
Insolation [W/m$^2$] & 50. & 0. \\
\hline
\end{tabular}
\end{center}
\end{table}

In the first part of the study \citep{Mills_Heygster2011c} a simple ice growth
model based on constant weather scenarios was described.
Here we apply the same model but with an extra step modelling brine expulsion
added.  These brine expulsion processes result from cooling of the ice sheet
in the upper layers:  because ice has a lower density than brine, as the ice
grows downward, brine in the upper layers will freeze generating pressure in
the remaining, more saline brine \citep{Tucker_etal1992}.

Since we are using two, constant-weather scenarios--see 
Table \ref{thermo_parm}--representing two possible
extremes, there is no need to create an iterative, time-based, prognostic
scheme, rather we take only ice thickness, $h$, in favour
of time, $t$, as the dependent variable in the bulk of the model.
The salinity profile is first modelled assuming no brine expulsion,
then there is a ``brine correction'' step, in which mass conservation
determines how much brine is lost based on the difference in temperature
between the ice layer at first formation and at the final ice thickness.

The following equation relates the ice surface temperature to
net heat flux:
\begin{equation}
h Q^* = k (T_\mathrm w - T_\mathrm s)
\label{thermo_model}
\end{equation}
where $T_{\mathrm w}$ is the water temperature which is assumed to be
constant at freezing (approximately -1.9$^\circ$ C at a water
salinity of 35 psu), $T_{\mathrm s}$ is surface temperature and $k$ is 
the thermal conductivity of the ice.
The net heat flux comprises the following components,
with functional dependencies supplied:
\begin{equation}
Q^* = Q_{\mathrm E} [e(T_{\mathrm s})] + Q_{\mathrm H} (T_{\mathrm s}) + Q_{\mathrm{SW}} (T_{\mathrm s}^4) + Q_{\mathrm{LW}}
\label{net_heat}
\end{equation}
The terms on the RHS are, from left to right:
latent heat, sensible heat, longwave and shortwave;
$e(T)$ is the saturation vapour pressure.
The first two terms are approximated with simple
parameterisations while the longwave flux is based
on the Stefan-Boltzmann law.
The shortwave flux is calculated primarily from
geometric considerations based on the position
of the Earth relative to the Sun.
The following inputs are required for the model:
surface-wind speed, -humidity, -air temperature
and -density (or pressure), cloud cover and
date and time or insolation.
\citep{Cox_Weeks1988,Drucker_etal2003,Yu_Lindsay2003}
Inputs can be supplied to give a 
picture of the general weather conditions.
For instance, fall freeze-up might be characterized
by relatively mild temperatures, low winds, high humidity,
high cloud cover and moderate insolation.
By contrast, a winter cold snap would be characterized by low
temperature, high winds, low humidity, clear conditions
and little to no insolation. See Table \ref{thermo_parm}.
Equations (\ref{thermo_model}) and (\ref{net_heat})
are solved with a numerical root-finding algorithm,
specifically bisection \citep{nr_inc2}.

The model can also be used as a crude ice growth model.
The rate of ice growth is:
\begin{equation}
g = \frac{Q^*}{L \rho_{\mathrm i}}
\label{ice_growth}
\end{equation}
where $L$ is the latent heat of fusion for water
and $\rho_{\mathrm i} \approx 0.917-1.403 \times 10^{-4} T$
is the density of pure ice, which is a function of temperature.
Empirical equations for determining the initial brine entrapment 
in sea ice have been derived by 
\citet{Cox_Weeks1988} and \citet{Nakawo_Sinha1981} and take the form:
\begin{equation}
S = S_0 f(g)
\label{salinity_from_growth}
\end{equation}
where $S_0$ is the salinity of the parent water and $f$ is
an empirical function of ice growth rate.

As shown in the previous report, if the salinities are left
untouched, even as they move further from the bottom of the ice
sheet, the model considerably over-estimates the bulk salinities.
Using conservation of mass, we can correct for this in each layer based on
the temperature differences between the thinner ice, when the layer
is still forming, and the thicker ice as more layers have been added
to the bottom. 
In the following, we go through all the relevant relationships
needed to form the brine expulsion-correction model.

While we will not need it, the time, $t$, can be written as a
 function of ice thickness:
\begin{equation}
t(h) = \int_0^h g^{-1}(h^\prime) \mathrm d h^\prime
\end{equation}
since the growth rate, $g$, is also a function of ice thickness.
Because we assume constant weather conditions, temperature for a given
point in the ice sheet is a monotonic function of ice age or thickness,
hence there is no reason to construct a dynamical time series.

Due to freezing-point depression, the salinity of brine inclusions,
$S_{\mathrm b}$, is a function of temperature only.  That is, greater salinity in
the brine will lower it's freezing point.  As the brine cools,
the water in it will freeze, increasing its salinity.
Thus it is maintained at its freezing point at all times.
We use the linear, piece-wise continuous model from 
\citet{Ulaby_etal1986}.

The total density at a given point in the ice is:
\begin{equation}
\rho=V_{\mathrm b} \rho_{\mathrm b} + (1-V_{\mathrm b}) \rho_{\mathrm i}
\label{total_density}
\end{equation}
where $V_{\mathrm b}$ is the relative brine volume, $\rho_{\mathrm b} \approx 1+0.0008 S_{\mathrm b}$
is the brine density \citep{Cox_Weeks1983}.  Note that this neglects any air pockets that are
usually included in the ice in addition to brine.  These are particularly
prevalent in older ice, but omitting them considerably simplifies the
analysis.

The ice salinity is further related to the brine salinity as follows:
\begin{equation}
S = S_{\mathrm b} \frac{\rho_{\mathrm b}}{\rho}
\label{ave_salinity}
\end{equation}
From equations (\ref{total_density}) and (\ref{ave_salinity}) we can
solve for the relative brine volume, $V_{\mathrm b}$:
\begin{equation}
V_{\mathrm b} = \frac{S \rho_{\mathrm i}}{S_{\mathrm b} \rho_{\mathrm b} - S (\rho_{\mathrm b} + \rho_{\mathrm i})}
\label{brine_volume}
\end{equation}
We assume that the ice is in thermodynamic equilibrium
and the thermal conductivity is roughly constant throughout.
Thus, the temperature of the ice sheet at depth, $z$, and thickness,
$h$, is given by interpolating between the surface temperature,
$T_{\mathrm s}$ and the water temperature, $T_{\mathrm w}$:
\begin{equation}
T (h,~z) = T_{\mathrm w} + [T_{\mathrm s}(h)-T_{\mathrm w}]z/h
\end{equation}
From this we can calculate the density of the ice
sheet as a function of ice thickness and ice depth: $\rho(h, ~z)$.
The relative change in volume, $\delta v$,
is given as:
\begin{equation}
\delta v = (\rho/\rho_0 - 1) 
\end{equation}
where $\rho_0(h)=\rho(h,~h)$ is the density of the ice layer as it is first forming.
Since the amount of mass must remain constant, and since we assume there are
no cavities in the ice, so must the volume remain constant, 
therefore we can determine how much material is ejected from the ice sheet.
We assume that only brine is ejected, provided there is enough, thus:
\begin{equation}
\delta m = \begin{cases} 
\rho_{\mathrm b} \delta v; & \delta v \le V_{\mathrm b} \\
\rho_{\mathrm b} V_{\mathrm b} + \rho_{\mathrm i} (\delta v - V_{\mathrm b}); & \delta v > V_{\mathrm b}
\end{cases}
\end{equation}
is the relative change in mass or relative mass loss.
It follows that the corrected brine volume is given as:
\begin{equation}
V_{\mathrm b}^{(\mathrm{new})} = \begin{cases}
V_{\mathrm b}-\delta v; & \delta v < V_{\mathrm b} \\
0; & \delta v \ge V_{\mathrm b}
\end{cases}
\end{equation}
from which we can solve for the corrected salinity using 
Equation (\ref{brine_volume}).

\section{Growth model results}

\label{growth_results}

\begin{figure}
\includegraphics[width=0.47\textwidth]{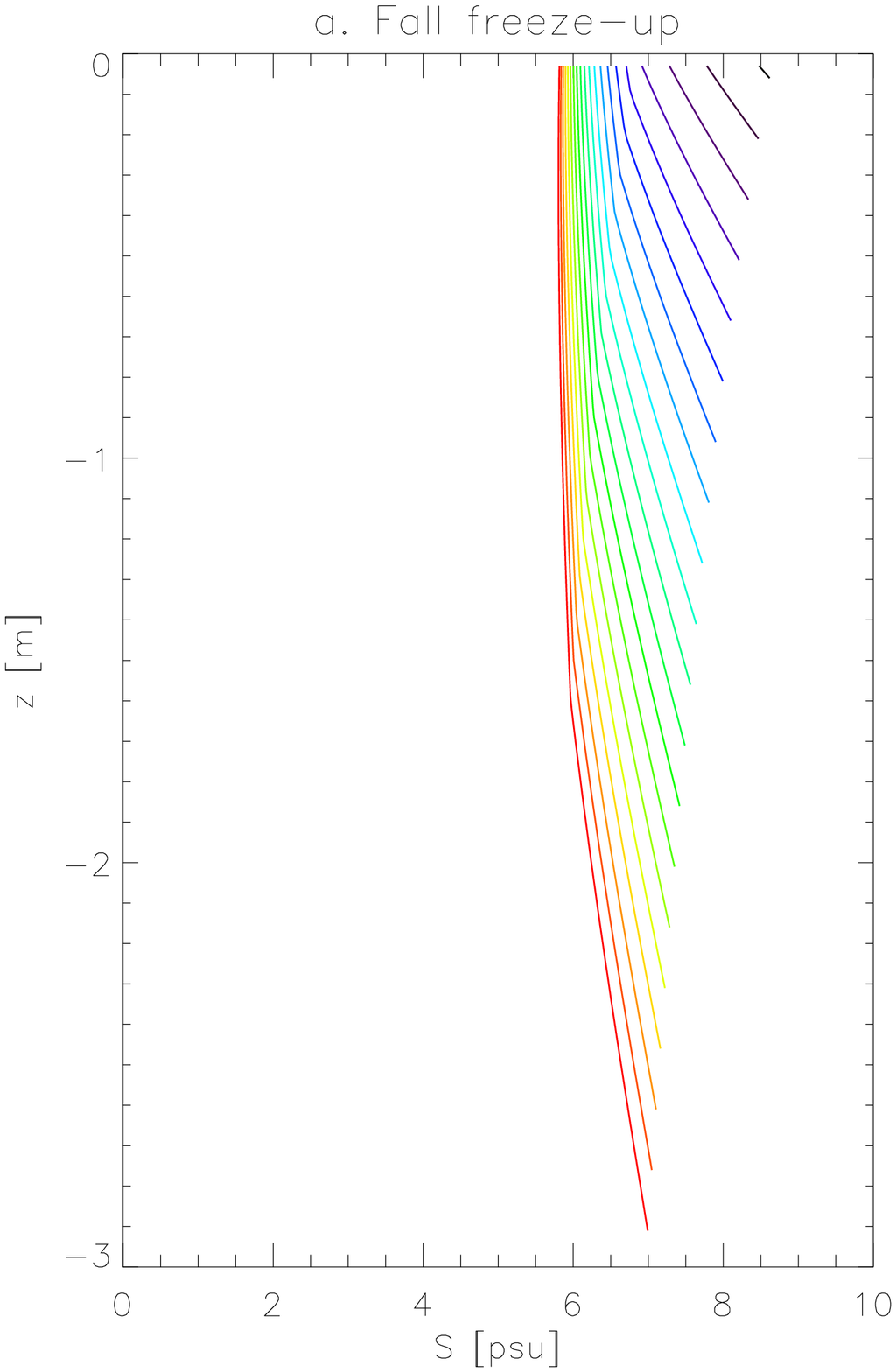}
\includegraphics[width=0.47\textwidth]{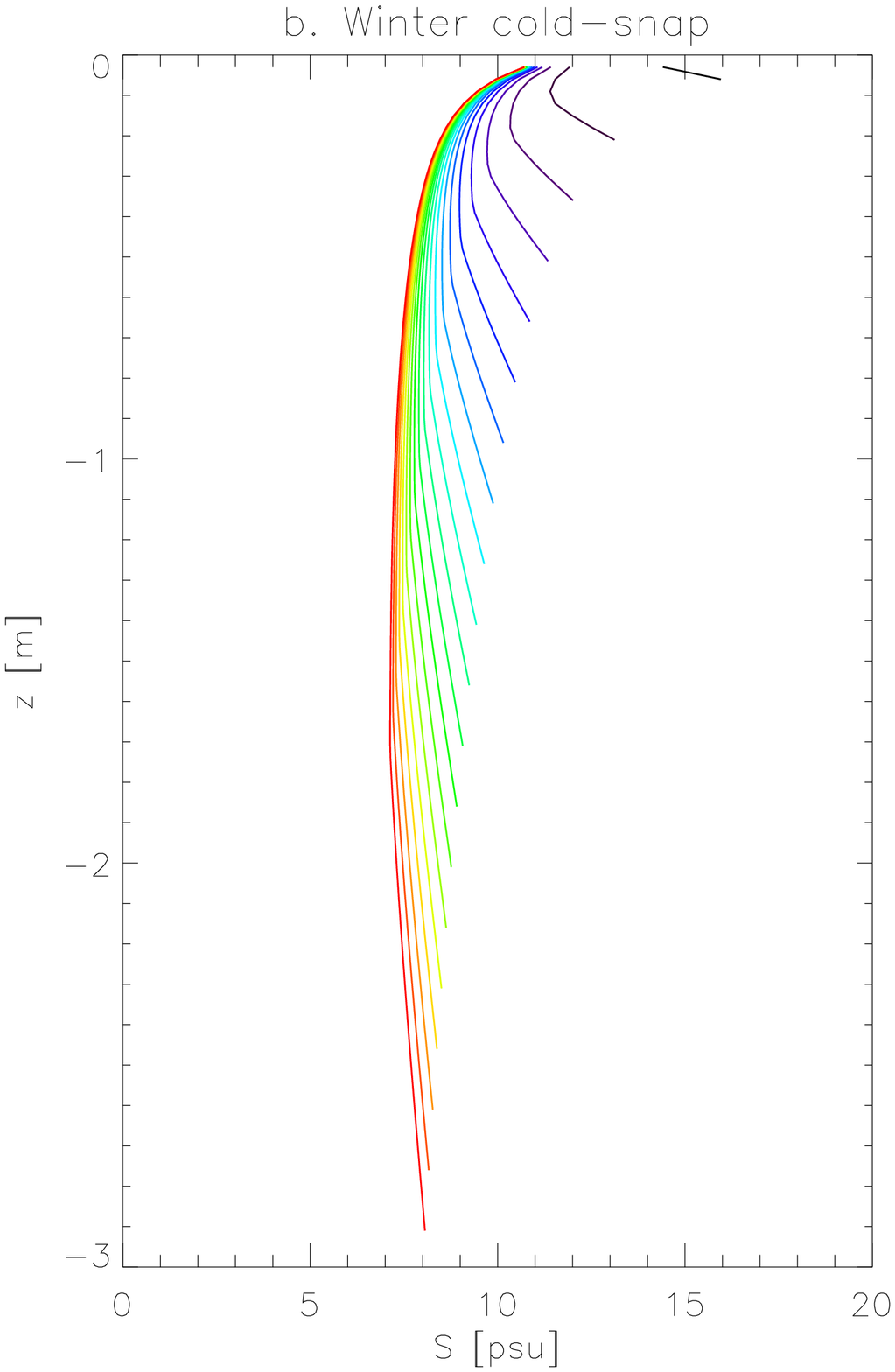}
\caption{Modelled salinity profiles for different ice thicknesses and
two weather scenarios: ``fall freeze-up'' and ``winter cold-snap'';
see Table \ref{thermo_parm}.}\label{sprof_mod}
\end{figure}

\begin{figure}
\begin{center}
\includegraphics[angle=90,width=0.9\textwidth]{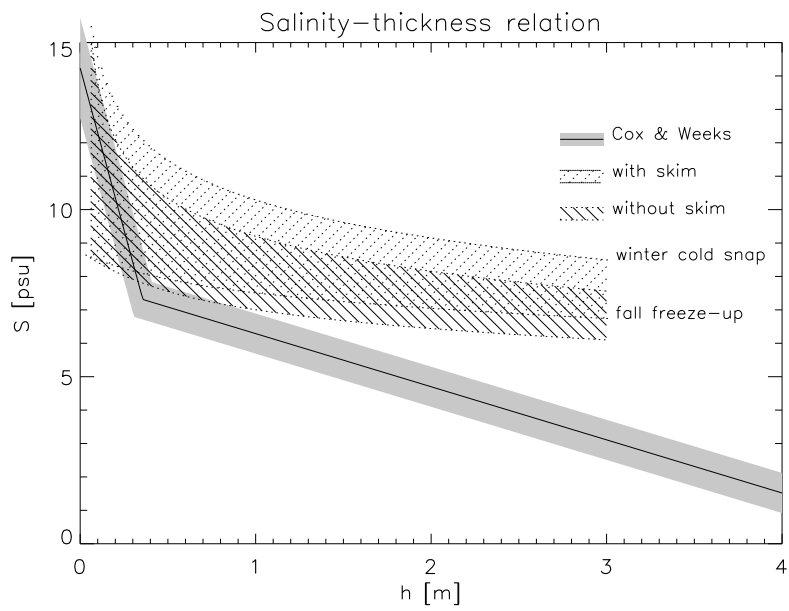}
\caption{Modelled versus empirical salinity-thickness relationships.
Solid shading shows the residuals for the \citet{Cox_Weeks1974} models.
The areas in between the two weather scenarios are hatched.
See text for an explanation of the difference between ``with skim''
and ``without skim.''}
\label{sh_rel}
\end{center}
\end{figure}

\begin{figure}
\begin{center}
\includegraphics[angle=90,width=0.9\textwidth]{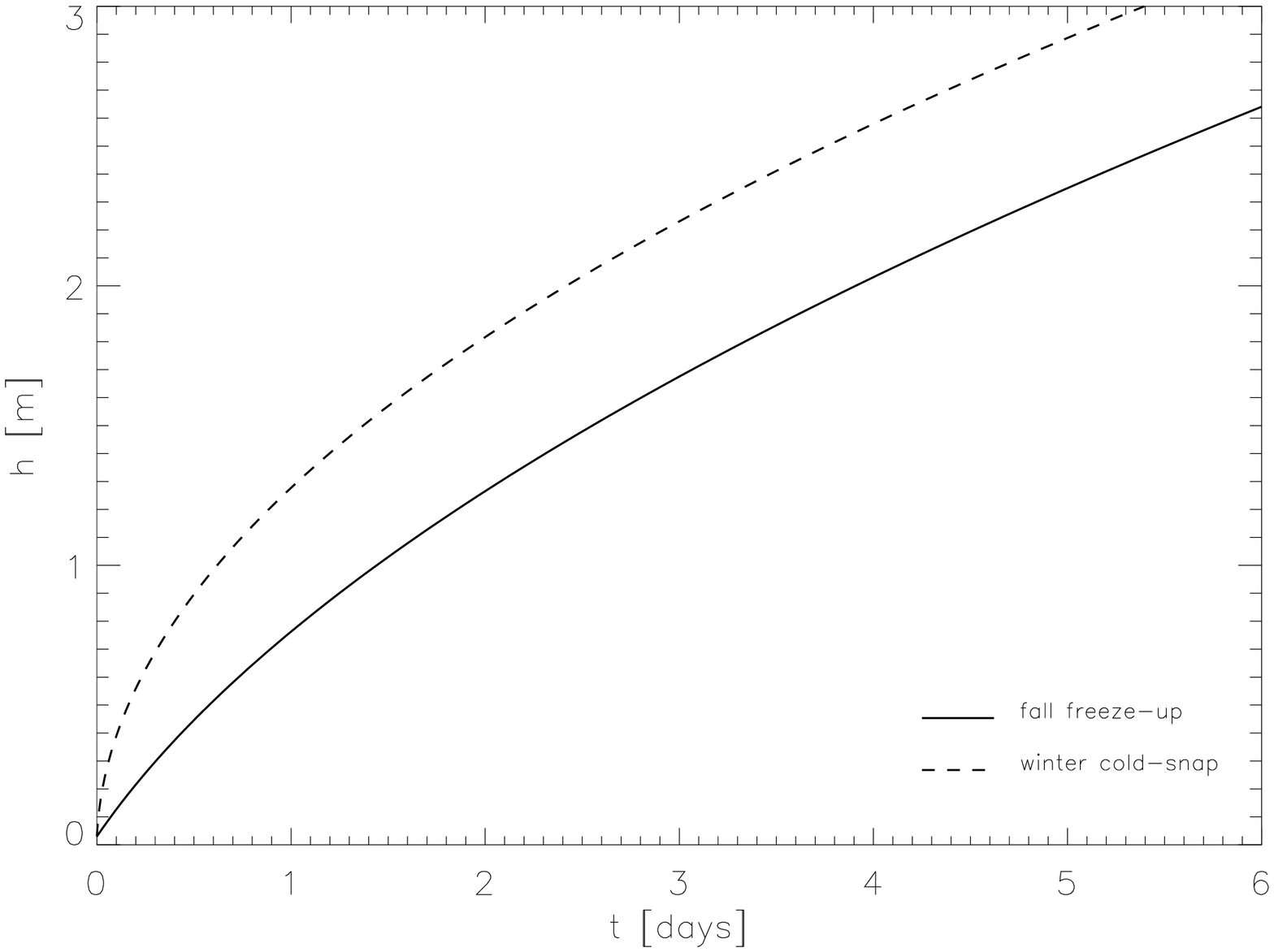}
\caption{Ice thickness as a function of time for the two contant-
weather scenarios.}\label{tvsh}
\end{center}
\end{figure}

Results for the ``fall freeze-up'' and ``winter cold-snap''
scenarios are shown in Figures \ref{sprof_mod}(a) and (b), respectively.
Qualitatively, at least, these salinity profiles are quite realistic:
thin-ice profiles for the cold-weather scenario show the characteristic,
`C'-shape typical of young ice \citep{Eicken1992}, 
while the thicker, older ice has a flatter profile \citep{Tucker_etal1992}.  
For the warm-weather scenario, the profiles are altogether flatter, 
which is typical of low-salinity ice \citep{Granskog_etal2006}.

Bulk salinity as a function of thickness are shown in Figure
\ref{sh_rel}, while thickness as a function of time is shown
in Figure \ref{tvsh}.
There are two versions of the growth model salinities: ``with skim''
and ``without skim'', both of which deserve some explanation.
The brine that has been ejected has to go somewhere.
In the first case, we assume that exactly half of it collects on top
of the ice sheet as ``skim'', while in the second case, we assume that
all of it is ejected downward, into the ocean.

\begin{figure}
\begin{center}
\includegraphics[angle=90,width=0.9\textwidth]{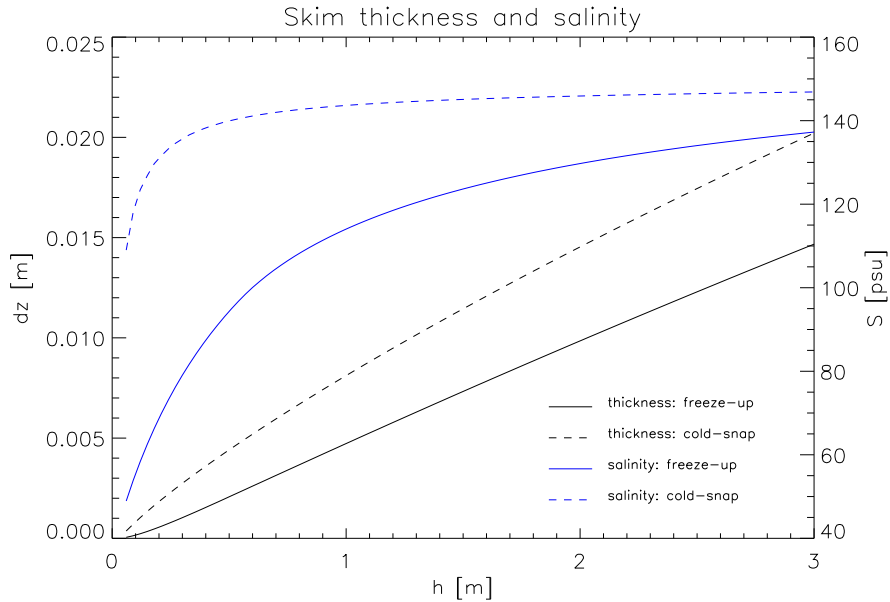}
\caption{Thickness and salinity of ``skim'' layer.}\label{skim}
\end{center}
\end{figure}

We need to calculate both the salinity of the skim, and its
thickness for further use in the emissivity model.
Following Equation (\ref{ave_salinity}), 
the lost salinity in each layer is:
\begin{equation}
S_\mathrm{lost} = \begin{cases} 
S_{\mathrm b} \rho_{\mathrm b} \delta v/\rho; & \delta v \le V_{\mathrm b} \\
S_{\mathrm b} \rho_{\mathrm b} V_{\mathrm b}/\rho; & \delta v > V_{\mathrm b}
\end{cases}
\end{equation}
The total mass loss is given as:
\begin{equation}
\delta m_\mathrm{total} = \int_0^h \delta m \mathrm d z
\end{equation} 
thus the salinity of the skim is calculated:
\begin{equation}
S_\mathrm{skim} = \frac{1}{\delta m_\mathrm{total}} \int_0^h S_\mathrm{lost} \delta m \mathrm d z
\end{equation}
Assuming that the skim is at the temperature of the top layer,
we can compute the density of the skim, $\rho_\mathrm{skim}$, from Equations 
(\ref{brine_volume}), then (\ref{total_density}), from which follows
its thickness:
\begin{equation}
\delta z_\mathrm{skim}=\frac{\delta m_\mathrm{total}}{2 \rho_\mathrm{skim}}
\end{equation}
Properties of the modelled skim layer are shown in Figure \ref{skim}.

Note that even if all the ejected brine is lost from the ice sheet,
the model still over-predicts the salinities in relation to the 
Cox and Weeks model, especially at deeper ice thicknesses.
The reason for the discrepency is that brine drainage has not been
included in the growth model: as the ice ages, brine will drain
through vertical channels.
Still, we hope that feeding these results to the emissivity models 
will nonetheless provide some insight.
For one thing, the salinity profile is different; also, there are more
emissivity variations in thinner ice, where the model is more accurate.

\section{Results}

\label{results}

\begin{figure}
\begin{center}
\includegraphics[angle=90,width=0.9\textwidth]{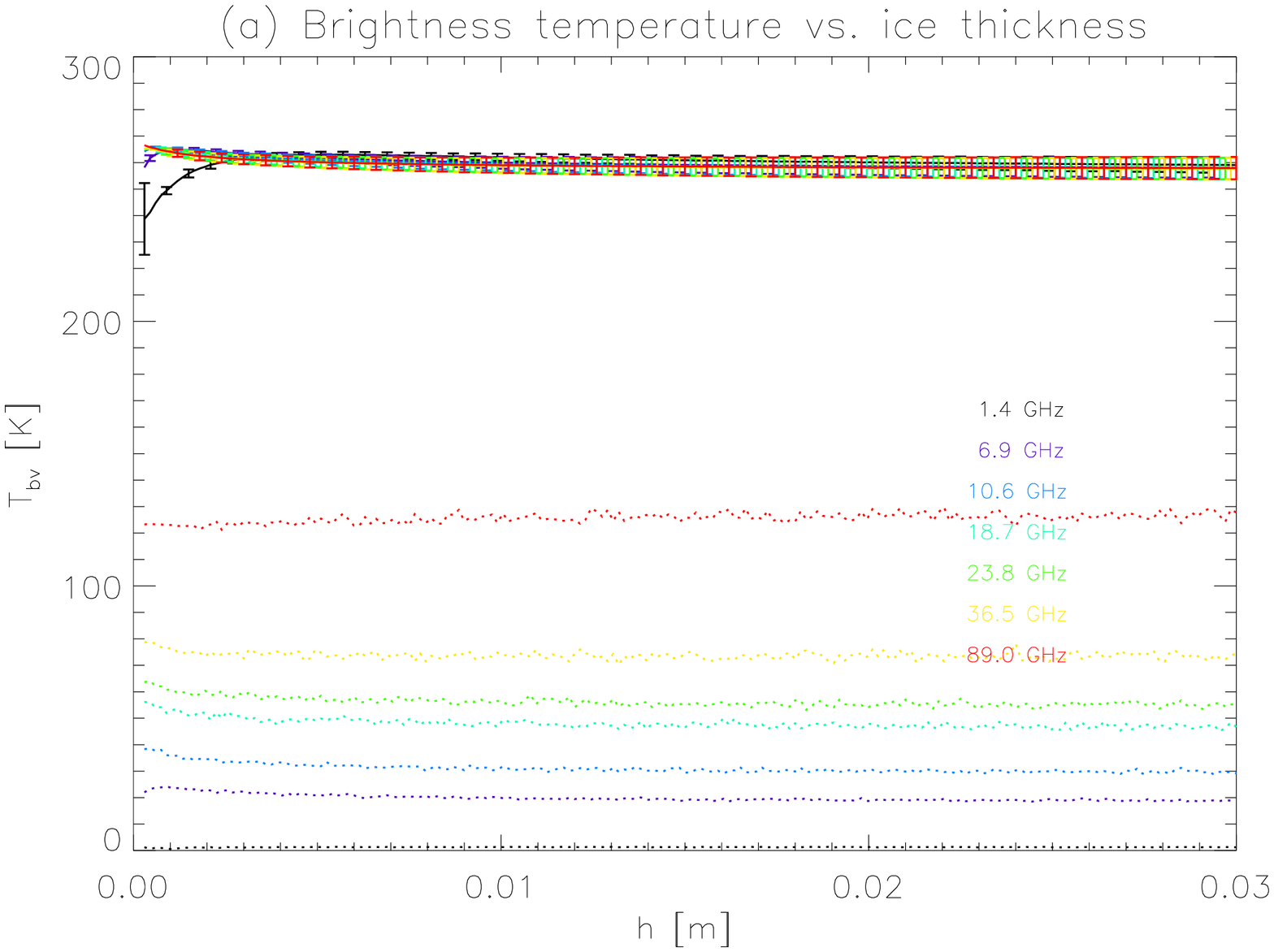}
\includegraphics[angle=90,width=0.9\textwidth]{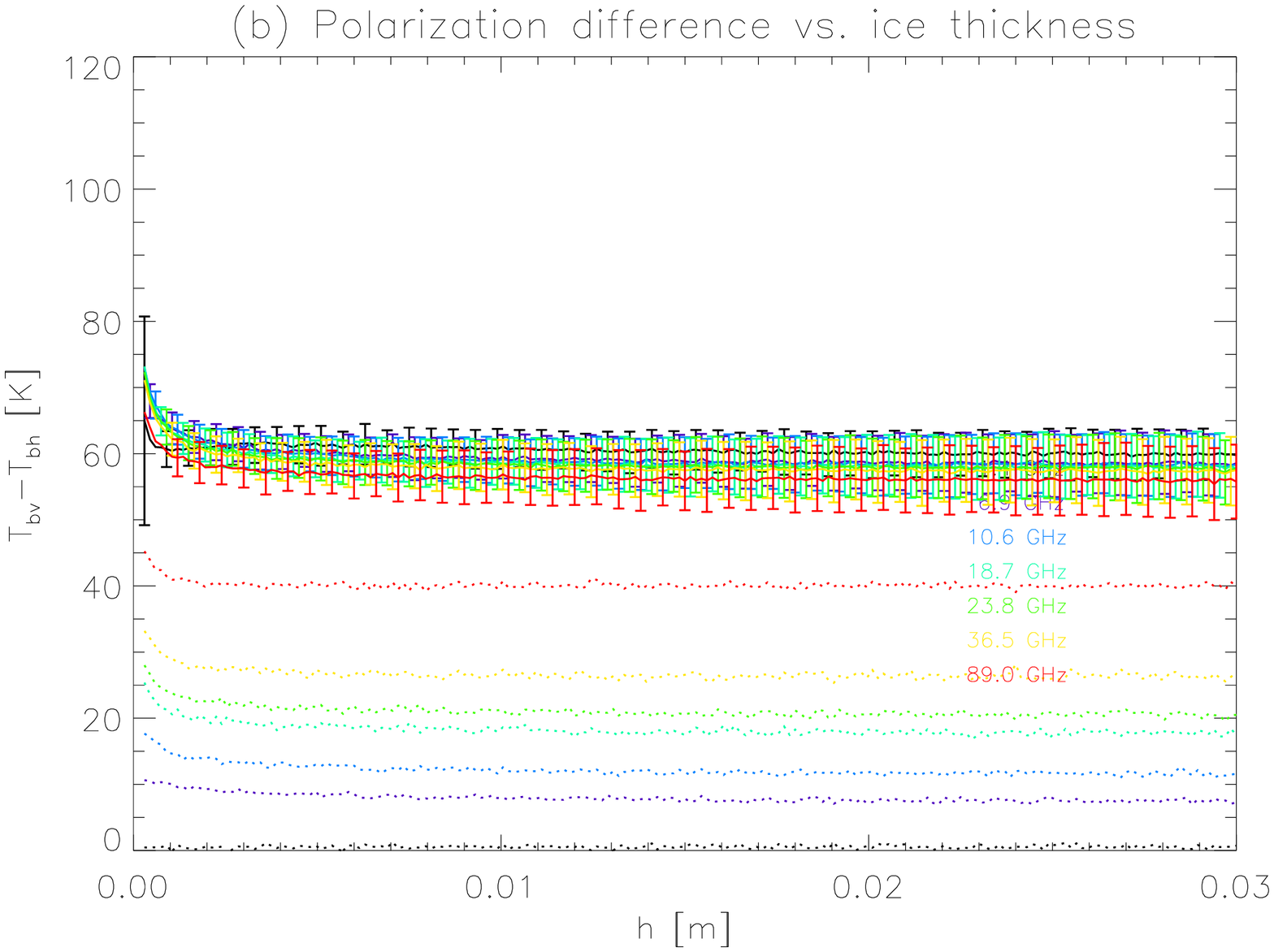}
\caption{Modelled brightness temperature as a function of ice thickness
based on modelled ice profiles, not including skim.
Solid lines show model results neglecting scattering, while
dotted lines show the computed scattering component.}
\label{tbvsh1a}
\end{center}
\end{figure}

\begin{figure}
\begin{center}
\includegraphics[angle=90,width=0.9\textwidth]{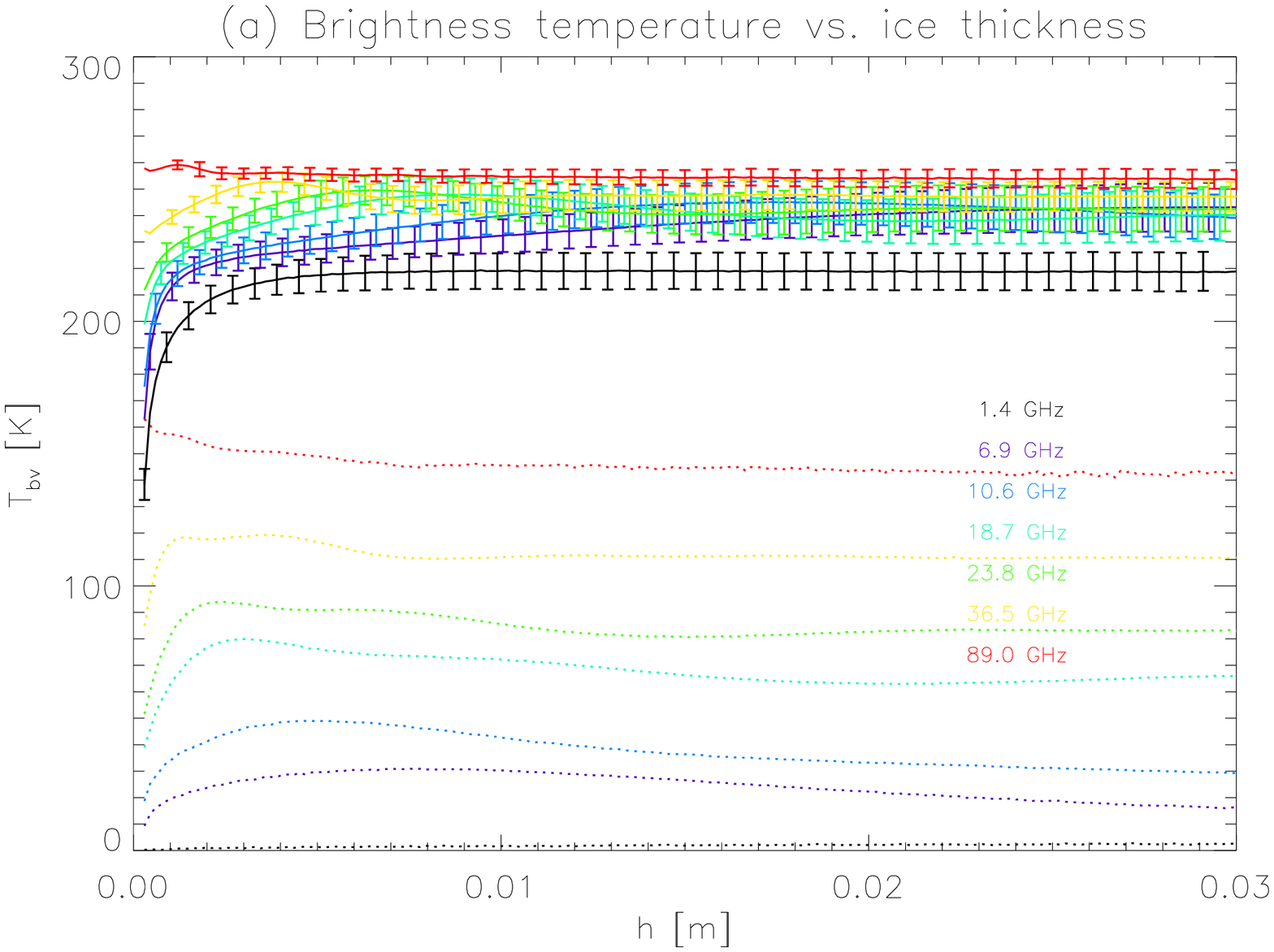}
\includegraphics[angle=90,width=0.9\textwidth]{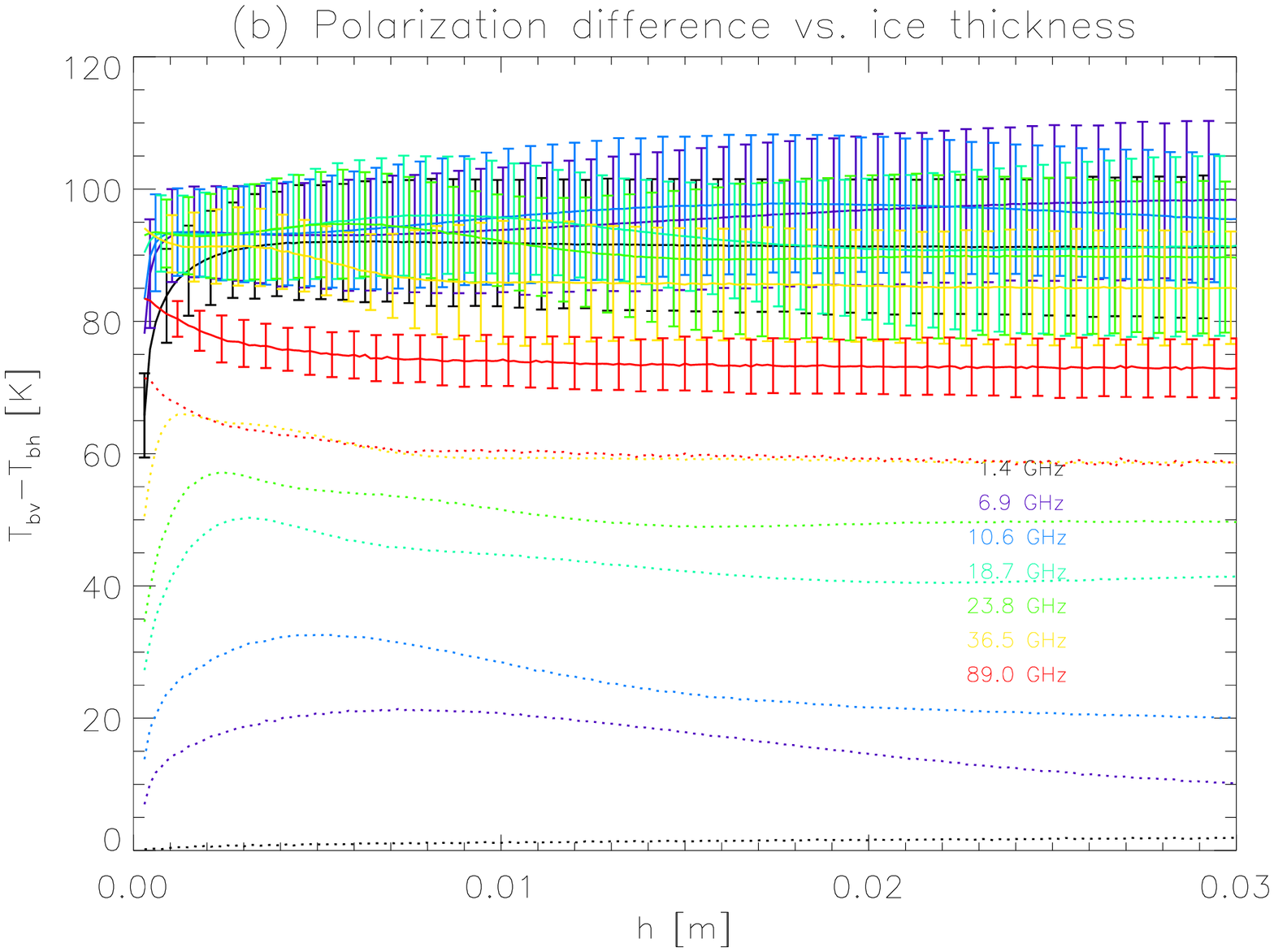}
\caption{Modelled brightness temperature as a function of ice thickness
based on modelled ice profiles, including skim.
Solid lines show model results neglecting scattering, while
dotted lines show the computed scattering component.}
\label{tbvsh1b}
\end{center}
\end{figure}

\begin{figure}
\begin{center}
\includegraphics[angle=90,width=0.9\textwidth]{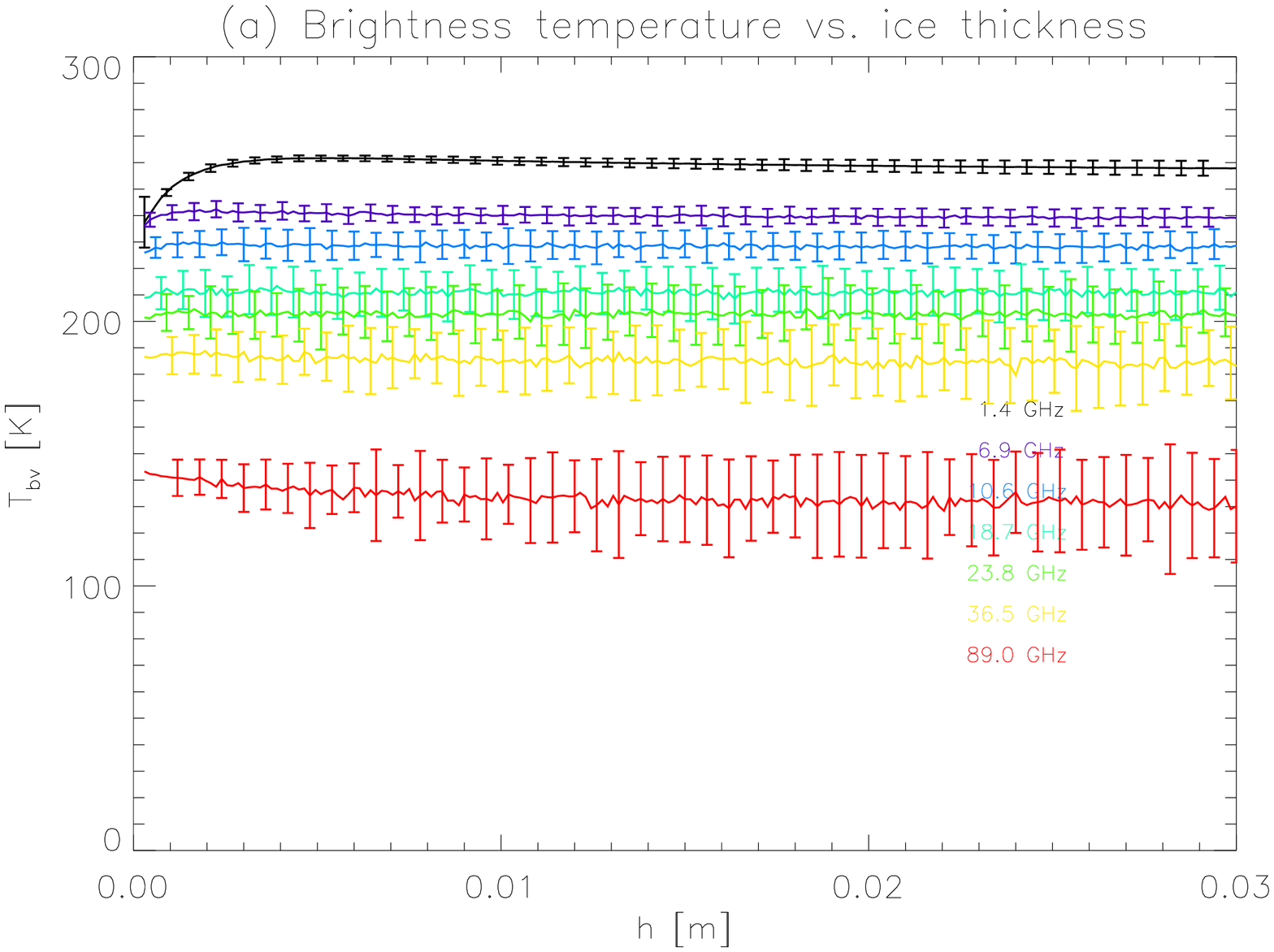}
\includegraphics[angle=90,width=0.9\textwidth]{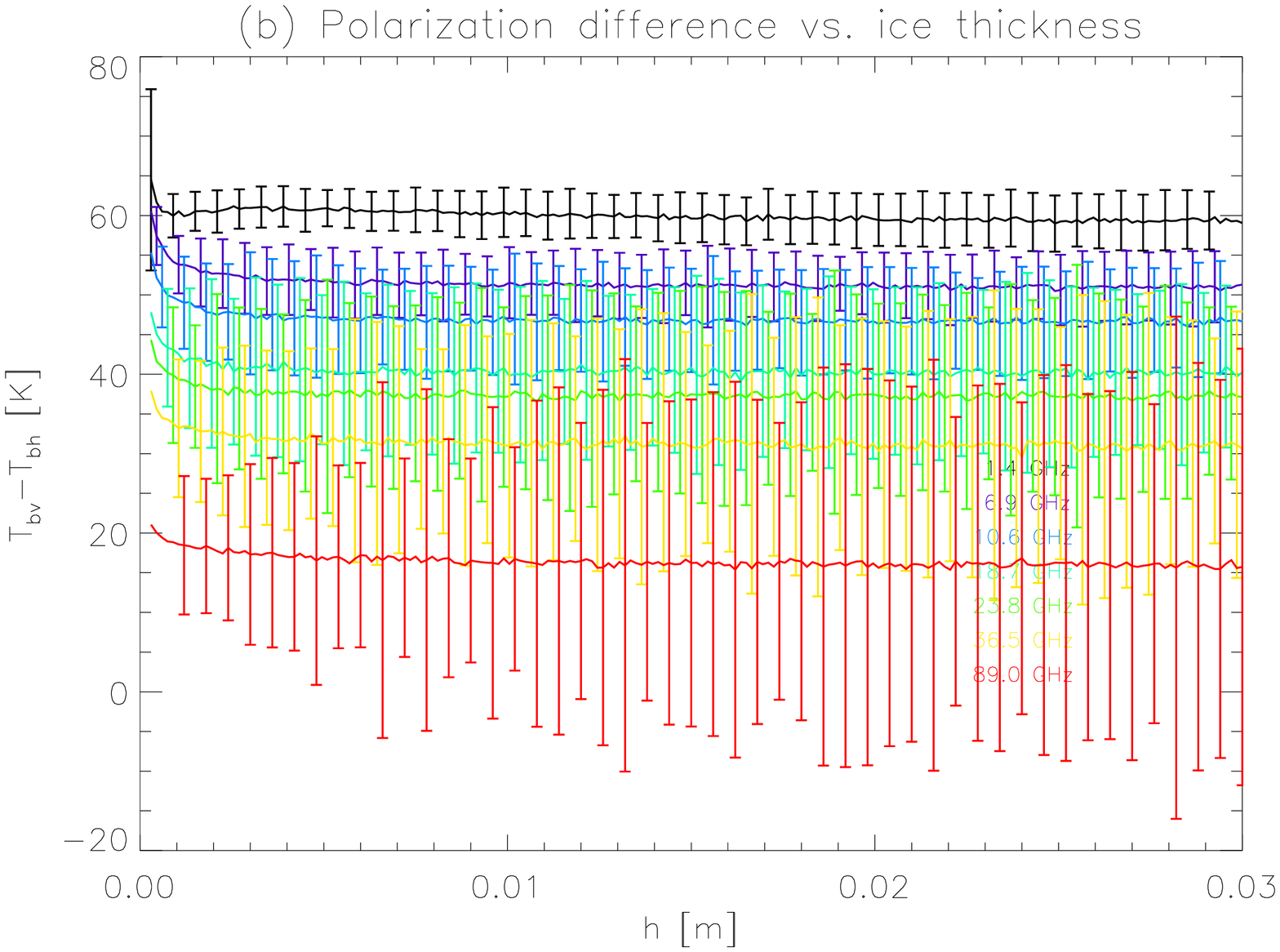}
\caption{Modelled brightness temperature as a function of ice thickness
based on modelled ice profiles, not including skim.
Scattering has been included.}
\label{tbvsh2a}
\end{center}
\end{figure}

\begin{figure}
\begin{center}
\includegraphics[angle=90,width=0.9\textwidth]{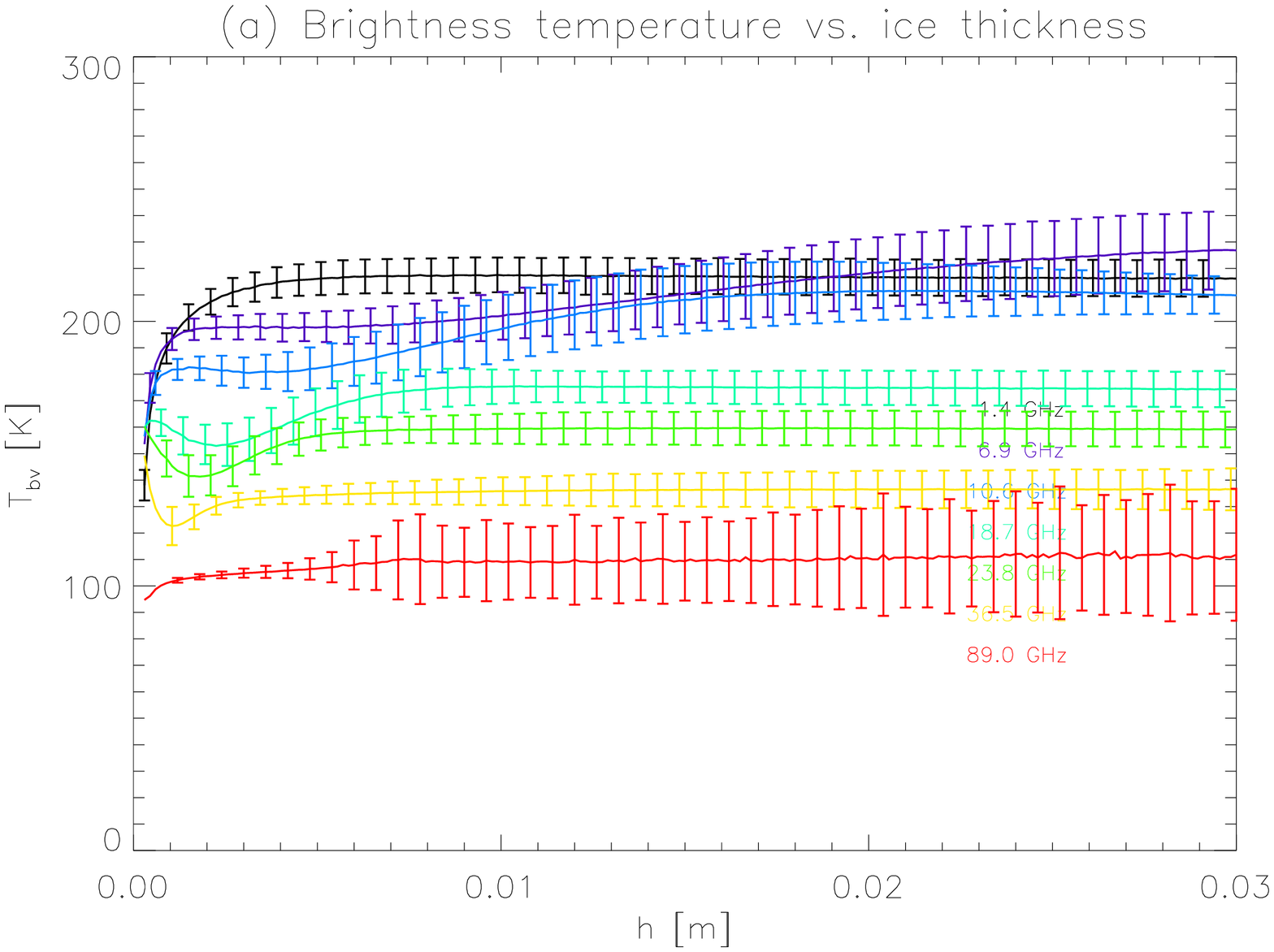}
\includegraphics[angle=90,width=0.9\textwidth]{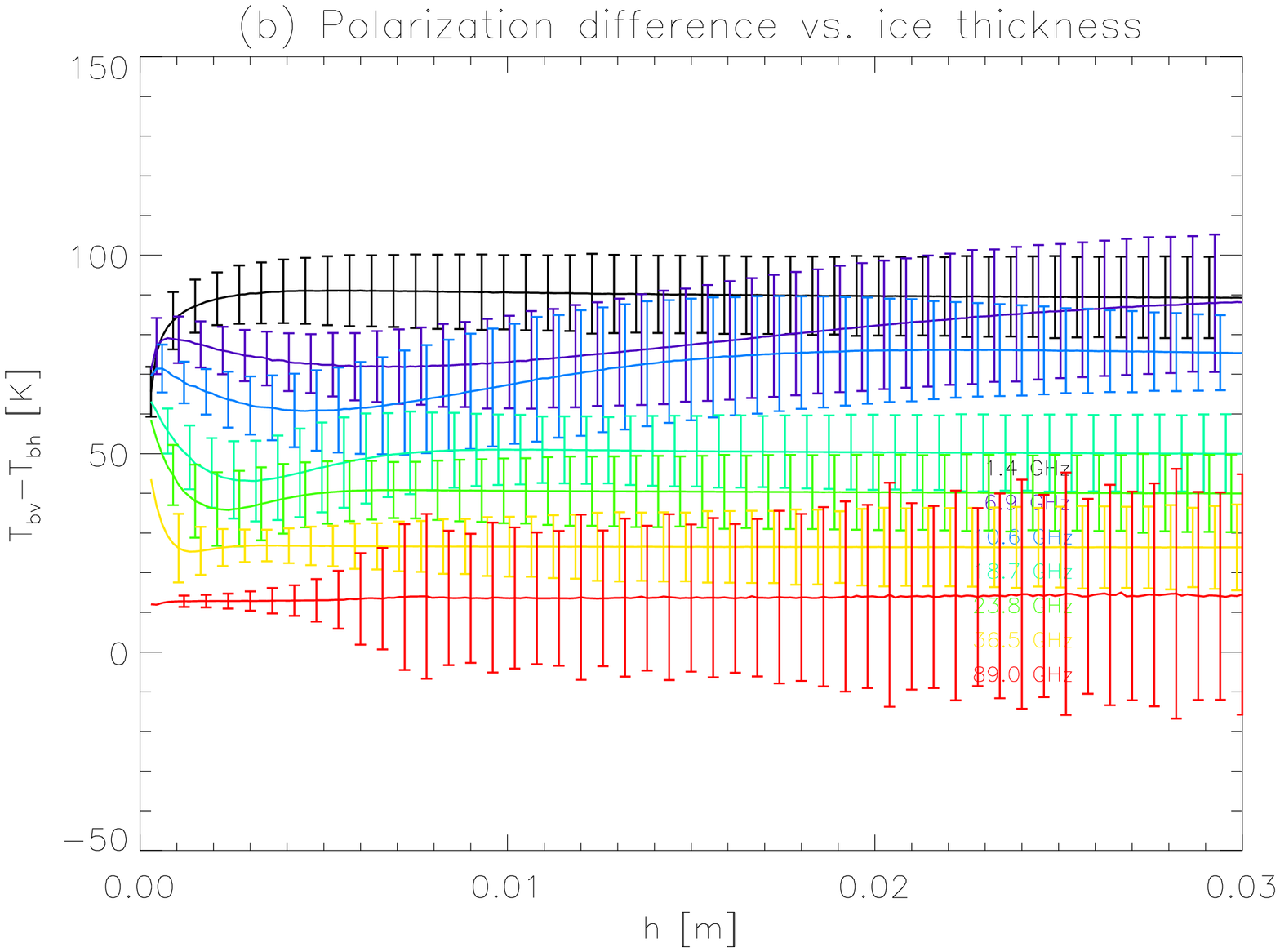}
\caption{Modelled brightness temperature as a function of ice thickness
based on modelled ice profiles, including skim.
Scattering has been included.}
\label{tbvsh2b}
\end{center}
\end{figure}

Here the final brightness temperature-thickness relations are presented.
In this second part, the error bounds have also been calculated using a
Monte Carlo or ``boot-strapping'' method \citep{nr_inc2}
by first varying the constant weather scenario between the two extremes
presented in Table \ref{thermo_parm} to produce 100 sets of profiles for
different ice thicknesses, then using the error bounds in the effective
permittivity model to vary the effective permittivities over 100 additional
trials, generating 10000 trials total.
Uniform deviates (``white'' noise) are used to vary the weather parameters
while normal deviates (Gaussian noise) are used to vary the permittivities. 

In the first set of graphs, Figure \ref{tbvsh1a}, 
neither the skim layer nor the scattering component has been
included in the emissivity calculations.
The scattering components, however, have been graphed alongside the
non-scattering components.
The second set of graphs, Figure \ref{tbvsh1b}, 
is the same as the first, except the skim layer has been included.
In the third set of graphs, Figure \ref{tbvsh2a}, 
the scattering and non-scattering components have been added together
for the simulations not including skim.
The fourth set of graphs, Figure \ref{tbvsh2b}, 
is the same as the third, except the skim layer has been included.

One of the most striking things about these results is how uniform the
brightness temperatures are in Figure \ref{tbvsh1a}, 
both for different ice thicknesses and for different frequencies.
Only the lowest frequency shows a slight dip at the thinnest ice
thicknesses because of transparency: the water ``shows through.''
Even with the addition of skim in Figure \ref{tbvsh1b},
the brightness temperatures are still surprisingly uniform.
Skim is one factor that is rarely included in 
either ice growth or ice emissivity models.
Because of its high liquid and high saline content, it will be
both highly reflective and highly absorbant, although because it is usually
a very thin layer, it may still be quite translucent.
Thus its presence will tend to lower the brightness temperature of the ice 
because it reflects the much cooler (as little as 3 K on clear days)
sky, as well as increase the polarization difference because of its
higher real permittivity.
These results show a fairly uniform depression/increase in  
brightness temperature/polarization difference,
but because the modelled skim does not drain as it would on real
ice, the actual tendency would apply only to new and thin ice.

The lower the frequency, the more both the real and the imaginary permittivity 
increase with brine volume
(see Figure 10 from Part I \citep{Mills_Heygster2011c}),
thus the effect of the
skim is more pronounced at lower frequencies.
Therefore, at lower frequencies, the characteristic lowered brightness
temperature and raised polarization difference of new and thin
ice is caused by the accumulation of skim.
At higher frequencies, this is caused by scattering.
The results here show that scattering increases the variability
of the signal especially at higher frequencies and especially
at higher ice thicknesses.
Note that skim also increases the variability, but more so
at lower frequencies.
In real ice we would expect the effect of scattering to be more pronounced
in thin rather than thick ice, and this does show up
somewhat in the model results, but only when skim is included.

\section{Summary and conclusions}

A simple thermodynamic model was used to simulate ice growth for
constant weather scenarios ranging between two extremes.
This model takes into account both heat conduction and brine expulsion
without the need to construct a dynamical time series, but neglects
brine drainage.
The resulting temperature and salinity profiles are fed to the
Microwave Emission Model of Layered Snowpack (MEMLS) to determine the
functional dependence of microwave radiance on ice thickness.
The resulting salinity profiles and consequent salinity-thickness
relation, as well as the simulated brightness-temperature-thickness
relations, are quite realistic.  The following observations may
be noted:
1. Without a layer of ``skim'' collected on top of the ice and
neglecting scattering, the brightness temperatures of the ice are surprisingly
uniform, both over different frequencies and different ice thicknesses.
Thus, variations in emissivity must be attributed to other, secondary 
factors which we describe below.

The presence of skim, a semi-liquid collection of brine and ice that
accumulates on top of ice sheets due to brine expulsion, is neglected in
most ice growth and emissivity models.
Here the skim has been modelled as simply the aggregate of exactly half
the brine that has been expelled from the ice sheet.
2. Because of its high opacity and high real permittivity,
the presence of skim will tend to lower the brightness temperature and increase
the polarization difference.
This effect is more pronounced at lower frequencies and at thinner
ice thickness, but the latter effect is not present because the growth
model neglects brine drainage.
3. Thus, the lower brightness temperature and higher polarization difference
characteristic of new and thin ice is primarily caused by the presence
of skim.

4. Finally, scattering will have a strong effect at higher frequencies.
5. Skim in particular will scatter strongly.
6. Thus, lower brightness temperature and higher polarization in thin
and new ice and lower brightness temperature and higher polarization
difference in general at higher frequencies is caused by scattering,
especially by the skim.  7. Both the presence of skim and scattering
will tend to increase the variability of the signal.

Both ice growth and the interaction of sea ice with electromagnatic
radiation are extremely complex processes with many different facets.
They are very difficult to model effectively.
Moreover, since all of: the microstructural properties, the macroscopic
configuration, the relative composition, will affect the emitted microwave
signal, and all these can change with the prevailing weather conditions,
ice growth and ice emissivity cannot be treated in isolation.
Figures 1 and 2 in Part I \citep{Mills_Heygster2011c} give some idea of the 
complexity of the problem.

We believe the contents of this and the previous report represent an
advance in the state-of-the-art, however there are still many limitations
in the current effort.
1. The ice growth model over-estimates salinity and the size of the skim
layer at high ice thicknesses because it neglects brine-drainage.
2. Scattering is too high and too variable at high ice thicknesses.
The scattering model is fairly crude and likely innaccurate, not least
of which because the correlation length are estimated by a simple,
ad-hoc assumption.
3. The complex permittivities do not take into account ice microstructural
properties, they do not have a comprehensive physical explanation
and they are not validated at higher frequencies.
As a first step, there are many new directions that the semi-empirical
permittivity model could be taken in.
A Debye relaxation curve for different brine volumes, for instance,
is suggested by the strong, simple dependence on frequency.

\bibliography{../final/sea_ice,../final/sft,../final/smos_final_extra,../final/smos_wp2.3a.bib,tstudy}

\begin{thebibliography}{}

\bibitem[\protect\astroncite{Cox and Weeks}{1983}]{Cox_Weeks1983}
Cox, G. and Weeks, W. (1983).
\newblock Equations for determining the gas and brine volumes in sea-ice
  samples.
\newblock {\em Journal of Glaciology}, 29(102):306--316.

\bibitem[\protect\astroncite{Cox and Weeks}{1988}]{Cox_Weeks1988}
Cox, G. and Weeks, W. (1988).
\newblock Numerical simulations of the profile properties of undeformed
  first-year sea ice during the growth season.
\newblock {\em Journal of Geophysical Research}, 93(C10):12499--12460.

\bibitem[\protect\astroncite{Cox and Weeks}{1974}]{Cox_Weeks1974}
Cox, G. F.~N. and Weeks, W.~F. (1974).
\newblock Salinity variations in sea ice.
\newblock {\em Journal of Glaciology}, 13(67):109--120.

\bibitem[\protect\astroncite{Drucker et~al.}{2003}]{Drucker_etal2003}
Drucker, R., Martin, S., and Moritz, R. (2003).
\newblock Observations of ice thickness and frazil ice in the {S}t. {L}awrence
  {I}sland polynya from satellite imagery, upward looking sonar, and
  salinity/temperature moorings.
\newblock {\em Journal of Geophysical Research}, 108(C5).

\bibitem[\protect\astroncite{Eicken}{1992}]{Eicken1992}
Eicken, H. (1992).
\newblock Salinity {P}rofiles of {A}ntarctic {S}ea ice: {F}ield {D}ata and
  {M}odel {R}esults.
\newblock {\em Journal of Geophysical Research}, 97(C10):15545--15557.

\bibitem[\protect\astroncite{Granskog et~al.}{2006}]{Granskog_etal2006}
Granskog, M., Kaartokallio, H., Kuosa, H., Thomas, D.~N., and Vainio, J.
  (2006).
\newblock Sea ice in the {B}altic {S}ea--{A} review.
\newblock {\em Estuarine, Coastal and Shelf Science}, 70:145--160.

\bibitem[\protect\astroncite{Hwang et~al.}{2007}]{Hwang_etal2007}
Hwang, B.~J., Ehn, J.~K., Barber, D.~G., Galley, R., and Grenfell, T.~C.
  (2007).
\newblock Investigations of newly formed sea ice in the {C}ape {B}athurst
  polynya: 2. {M}icrowave emission.
\newblock {\em Journal of Geophysical Research}, 112(C05003).

\bibitem[\protect\astroncite{Kwok et~al.}{2007}]{Kwok_etal2007}
Kwok, R., Comiso, J., Martin, S., and Drucker, R. (2007).
\newblock Ross {S}ea polynyas: Response of ice concentration retrievals to
  large areas of thin ice.
\newblock {\em Journal of Geophysical Research}, 112(C12012).

\bibitem[\protect\astroncite{Martin et~al.}{2004}]{Martin_etal2004}
Martin, S., Drucker, R., Kwok, R., and Holt, B. (2004).
\newblock Estimation of the thin ice thickness and heat flux for the {C}hukchi
  {S}ea {A}laskan coast polynya from {S}pecial {S}ensor {M}icrowave {I}mager
  data, 1990-2001.
\newblock {\em Journal of Geophysical Research}, 109(C10012).

\bibitem[\protect\astroncite{Mills and Heygster}{2011}]{Mills_Heygster2011c}
Mills, P. and Heygster, G. (2011).
\newblock Sea ice emissivity as a function of ice thickenss: computed curves
  for {AMSR-E} and {SMOS} (frequencies from 1.4 to 89 {GH}z).
\newblock Technical Report DFG project HE-1746-15, University of Bremen.

\bibitem[\protect\astroncite{Nakawo and Sinha}{1981}]{Nakawo_Sinha1981}
Nakawo, M. and Sinha, N.~K. (1981).
\newblock Growth rate and salinity profile of first-year sea ice in the high
  arctic.
\newblock {\em Journal of Glaciology}, 27(96):315--330.

\bibitem[\protect\astroncite{Naoki et~al.}{2008}]{Naoki_etal2008}
Naoki, K., Ukita, J., Nishio, F., Nakayama, M., Comiso, J.~C., and Gasiewski,
  A. (2008).
\newblock Thin sea ice thickness as inferred from passive microwave and in situ
  observations.
\newblock {\em Journal of Geophysical Research}, 113(C02S16).

\bibitem[\protect\astroncite{Press et~al.}{1992}]{nr_inc2}
Press, W.~H., Teukolsky, S.~A., Vetterling, W.~T., and Flannery, B.~P. (1992).
\newblock {\em Numerical Recipes in C}.
\newblock Cambridge University Press, second edition.

\bibitem[\protect\astroncite{Tucker et~al.}{1992}]{Tucker_etal1992}
Tucker, W.~B., Perovich, D.~K., Gow, A.~J., Weeks, W.~F., and Drinkwater, M.~R.
  (1992).
\newblock Physical {P}roperties of {S}ea {I}ce {R}elevant to {R}emote
  {S}ensing.
\newblock In {\em Microwave Remote Sensing of Sea Ice}, number~68 in
  Geophysical Monographs, chapter~2, pages 9--28. American Geophysical Union.

\bibitem[\protect\astroncite{Ulaby et~al.}{1986}]{Ulaby_etal1986}
Ulaby, F.~T., Moore, R.~K., and Fung, A.~K., editors (1986).
\newblock {\em Microwave Remote Sensing: Active and Passive, Volume III, From
  Theory to Applications}.
\newblock Artech House, Norwood, MA.

\bibitem[\protect\astroncite{Yu and Lindsay}{2003}]{Yu_Lindsay2003}
Yu, Y. and Lindsay, R.~W. (2003).
\newblock Comparison of thin ice thickness derived from {RADARSAT}
  {G}eophysical {P}rocessor {S}ystem and {A}dvanced {V}ery {H}igh {R}esolution
  {R}adiometer data sets.
\newblock {\em Journal of Geophysical Research}, 108(C12).

\end{thebibliography}

\end{document}